# Ecology theory disentangles microbial dichotomies

Luciana L. Couso[1,*], Alfonso Soler-Bistué[2,*], Ariel A. Aptekmann[3], Ignacio E. Sánchez[3]

[1] Universidad de Buenos Aires. Facultad de Agronomía. Cátedra de Genética. Buenos Aires, Argentina
[2] Instituto de Investigaciones Biotecnológicas Dr. Rodolfo A. Ugalde, CONICET, Universidad Nacional de San Martín, Argentina.
[3] Universidad de Buenos Aires. Consejo Nacional de Investigaciones Científicas y Técnicas. Instituto de Química Biológica de la Facultad de Ciencias Exactas y Naturales (IQUIBICEN). Facultad de Ciencias Exactas y Naturales. Laboratorio de Fisiología de Proteínas. Buenos Aires, Argentina.

* These authors contributed equally to this work

Correspondence may be addressed to LLC (lcouso@agro.uba.ar), ASB (asoler@iib.unsam.edu.ar) and IES (isanchez@qb.fcen.uba.ar)

Microbes are often discussed in terms of dichotomies such as copiotrophic/oligotrophic and fast/slow-growing microbes, defined using the characterisation of microbial growth in isolated cultures. The dichotomies are usually qualitative and/or study-specific, sometimes precluding clear-cut results interpretation. We are able to interpret microbial dichotomies as life history strategies by combining ecology theory with Monod curves, a classical laboratory tool of bacterial physiology. Monod curves relate the specific growth rate of a microbe with the concentration of a limiting nutrient, and provide quantities that directly correspond to key ecological parameters in McArthur and Wilson's r/K selection theory, Tilman's resource competition and community structure theory and Grime's triangle of life strategies. The resulting model allows us to reconcile the copiotrophic/oligotrophic and fast/slow-growing dichotomies as different subsamples of a life history strategy triangle that also includes r/K strategists. We analyzed some ecological context by considering the known viable carbon sources for heterotrophic microbes in the framework of community structure theory. This partly explains the microbial diversity observed using metagenomics. In sum, ecology theory in combination with Monod curves can be a unifying quantitative framework for the study of natural microbial communities, calling for the integration of modern laboratory and field experiments.

## Introduction

Microbes (bacteria, archaea and unicellular eukaryotes) constitute close to 20% of the known species and over 50% of the known phylogenetic diversity (1, 2). Current experimental techniques such as metagenomics often find evidence for thousands of distinct microbial genomes in a single natural environment (3). Many of these genomes also possess an active gene expression machinery (4) (Figure 1, left). A simplified view of microbial diversity classifies organisms with reference to two archetypes, usually defined using the physiological and genomic traits of a small number of model microbes growing in isolated cultures (5). Such laboratory-focused copiotrophic/oligotrophic dichotomy is closely related to Winogradsky's ecological zymogenic/autochthonous dichotomy (5, 6) and was first defined by the ability to thrive at high and low concentrations of a growth-limiting nutrient, respectively (7). This classification can be further split into exclusive copiotrophs/oligotrophs, where oligotrophs only grow at low nutrient concentrations and copiotrophs only grow at high nutrient concentrations, and coexisting copiotrophs/oligotrophs, which can both thrive in nutrient-poor and nutrient-rich media (5, 8). Here, we focus on coexisting copiotrophs/oligotrophs. Coexisting oligotrophs (oligotrophs from now on) are expected to present higher-affinity nutrient uptake systems, higher efficiency of resource use at the expense of a lower growth rate, a larger surface-to-volume ratio and a low-cost, slower protein expression machinery relative to coexisting copiotrophs (copiotrophs from now on) (7–12). The genomic traits reported by different studies to correlate with copiotrophs/oligotrophs are only partially overlapping (5, 10–13) and thus this dichotomy remains to be fully understood (14). On the other hand, the fast versus slow-growing microbes dichotomy was defined in terms of their specific growth rates, regardless of nutrient availability, and related to the optimization gene transcription and translation (5, 15, 16). Still, optimization of translation is a poor predictor of the maximum specific growth rate of slow-growing microbes (15). In addition, the relationship between the copiotrophic/oligotrophic and the fast versus slow-growing dichotomies is unclear (5). In this work, we aim at clarifying the similarities and differences between these dichotomies using Monod curves and ecological theory.

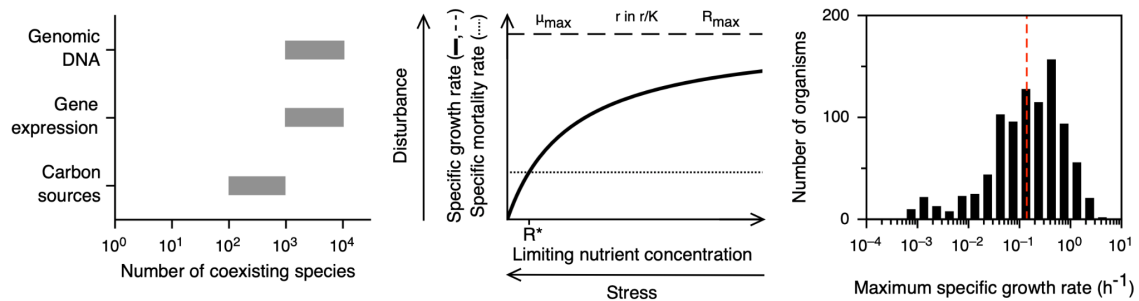

*Figure 1. Microbial diversity in relation to Monod curves. Left: Typical number of coexisting species in a natural microbial community as estimated from metagenomics experiments analyzing genomic DNA (3), the number of species with an active gene expression machinery in metatranscriptomics experiments (4) and the number of different carbon sources on which heterotrophic prokaryotes may grow (1, 17, 18). Middle: Specific growth rate of a microbe during balanced growth of an isolated culture, µ, as a function of the concentration of a growth-limiting substrate S (Monod curve, solid line). The dashed line indicates $\mu_{max}$, the maximum specific growth rate of this microbe during balanced growth. The dotted line indicates the specific mortality rate, which we assume to be independent of limiting nutrient concentration. The population is viable if the Monod curve is above the line for the specific mortality rate. At the point where the two curves cross, growth and mortality balance each other, leading to a*

*stable number of individuals. The corresponding limiting nutrient concentration is denoted as R\*. The position along the x axis is inversely proportional to stress in Grime's CSR model, while the position of the specific mortality rate along the y axis is proportional to disturbance in Grime's CSR model. Right: distribution of maximum specific growth rates in bacteria and archaea (1). The vertical red line corresponds to a doubling time of 5 hours, which is used in (15) as an empirical separation between fast and slow-growing microbes.*

**Results**

**Copiotrophic/oligotrophic and fast versus slow-growing microbes in relation to Monod curves**

We can analyze the copiotrophic/oligotrophic and fast versus slow-growing microbes dichotomies in terms of the relationship between the specific growth rate of a microbe in isolated culture and the concentration of a growth-limiting nutrient (19), i.e., their Monod curves (5). Such curves usually follow the empirical Monod equation (Figure 1, middle):

$$\mu = \frac{\mu_{max} \cdot [S]}{[S] + K_S} \qquad [1]$$

Where μ is the specific growth rate (sometimes called relative growth rate) of a microbe during balanced growth of a pure, axenic culture, [S] is the concentration of a growth-limiting substrate S, $\mu_{max}$ is the maximum specific growth rate of this microbe during balanced growth and $K_S$ is the value of [S] for which μ is half of the maximum (Table 1). As long as the Monod equation holds, μ approaches $\mu_{max}$ asymptotically at infinite values of S, i.e., when there is no limitation for growth. The specific growth rate of a microbe is normalized relative to biomass in the culture and can be used to compare the growth of different microbes. The distribution of the maximum specific growth rates in bacteria and archaea spans nearly four orders of magnitude (1) (Figure 1, right), suggesting that $\mu_{max}$ is relevant to the characterization of microbial diversity. In a sparsely populated environment, the specific growth rate of a microbe is a useful laboratory estimation of its fitness (20). Other important parameters for the characterization of microbial growth are the specific mortality rate and the yield for conversion of nutrients to biomass (Table 1). The specific mortality rate, which to a first approximation we assume to be independent of limiting nutrient concentration, can be readily incorporated in a Monod graph (Figure 1, middle). The yield can be measured using standard laboratory techniques (10), but it can not be extracted from Monod curves, which are measured during balanced microbial growth when the culture is far from saturation. Microbes with a higher nutrient-to-biomass yield are better suited to sustain higher population densities.

Copiotrophs have a higher maximum growth rate than oligotrophs but attain it at higher concentrations of growth-limiting nutrients (5, 8, 9). In other words, copiotrophs present a higher $\mu_{max}$ and a higher $K_S$ (the [S] for which μ equals $\mu_{max}$/2) than oligotrophs, leading to intersecting Monod curves (Figure 2, left, and Table 2). At nutrient concentrations higher than the crossing of the two Monod curves, copiotrophs perform better, while at nutrient concentrations below the crossing point, oligotrophs prevail. Coexistence of copiotrophs and oligotrophs in nature could be explained by a fluctuating nutrient supply that is alternatively higher and lower than the crossing point. A trade-off between growth rate and yield for coexisting copiotrophs/oligotrophs has been proposed from the experimental study of eight microbes, but not tested extensively (10). On the other hand, fast-growing microbes present a higher specific growth rate than slow-growing microbes at all experimentally accessible concentrations of the growth-limiting nutrient (5, 15) (Figure 2, right, and Table 2). Thus, the coexistence of fast and slow-growing microbes in nature can not be explained from Monod curves. A possible rationalization is that slower-growing microbes may have a higher yield for conversion of nutrients into biomass and thus prevail in high density environments. Alternatively, dormant forms such as spores, persister or viable but nonculturable cells might also enable coexistence.

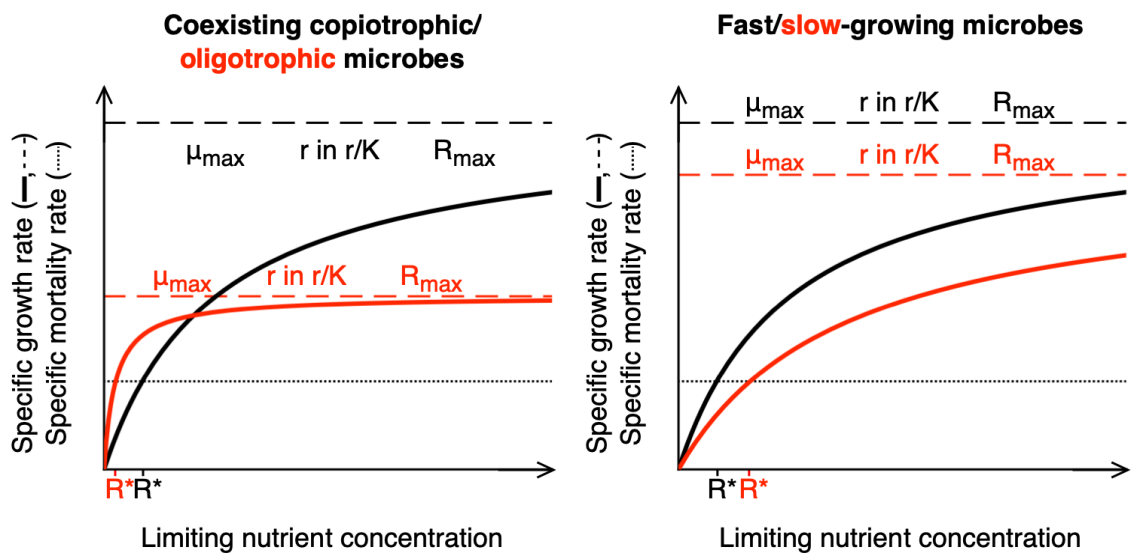

***Figure 2. Microbial dichotomies as pairs of Monod curves.*** *Specific growth rate of a microbe during balanced growth of an isolated culture as a function of the concentration of a growth-limiting substrate S (solid line). The dotted line indicates the specific mortality rate, i.e., the minimal specific growth rate required to maintain a stable number of individuals, which we assume to be independent of limiting nutrient concentration. The corresponding limiting nutrient concentrations are denoted as R\*. The dashed lines indicate the maximum specific growth rate of a microbe during balanced growth. Left: Monod curves of copiotrophs (black line) and oligotrophs (red line), measured for each microbe in isolation. Right: Monod curves of fast (black line) versus slow-growing (red line) microbes, measured for each microbe in isolation.*

We will now relate Monod curves and their main parameters with three ecological theories relevant to microbial diversity and show how these relationships can help us rationalize resource competition and life strategies in relation to copiotrophic/oligotrophic and fast versus slow-growing microbes.

**Relation to r/K selection theory**

Wilson and McArthur's theory of island biogeography (21, 22) describes the successional observation of species during colonization of a habitat, from early conditions of low number of individuals and readily available resources to later conditions of sustained crowding and intense competition for resources. This is similar to Winogradsky's successional observation of zymogenic and autochthonous species after addition of organic matter to soil (6), which contributed to the formulation of the copiotrophic/oligotrophic dichotomy (5). Wilson and McArthur build a quantitative description of this time course around two parameters: the maximum specific growth rate of a species (r) and the density of individuals that a given environment can support at equilibrium in isolation (the carrying capacity K). It is widely assumed that there is a trade-off between r and K, so that evolution can not maximize both parameters. This defines an r/K continuum, with two archetypal life strategies at the extremes. r-strategists grow fast to a low equilibrium density and thrive in uncrowded environments where perturbations (e.g., sporadic storms, desiccation, or temperature extremes) affect essentially the same proportion of individuals at all densities. Conversely, K-strategists grow slowly to a high equilibrium density and thrive in crowded environments where populations are limited by density-dependent controls, such as food supply, toxic metabolites, or predation. During habitat colonization, r-strategists thrive first and K-strategists prevail at longer times.

The similarities between r/K-strategists and copiotrophs/oligotrophs have been already discussed (23). Overall, the traits commonly associated with r/K strategists and copiotrophic/oligotrophic microbes overlap to some degree (11, 24). Recent experiments directly addressed this question by studying changes in the gene pool of soil microbial communities induced in two ways (12). Changes in the functional repertoire of a bacterial community along a nutrient gradient were assumed to arise from changes in the relative populations of copiotrophs and oligotrophs in the culture (Figure 2, left), while changes in the gene pool along a density gradient were assumed to arise from changes in the relative populations of r and K strategists in the culture (12). This procedure allowed to assign axes of functional gene variation related to the r/K categories and the copiotrophic/oligotrophic categories. These axes were markedly different, suggesting that these two classifications should not be merged by default (12). In sum, although r/K strategies resemble the nutrient-based copiotrophic/oligotrophic split, a more nuanced description might be within our reach by embracing an ecology-rooted, successional view of microbe classification.

*Table 1. Equivalence between parameters of Monod curves and other laboratory experiments and r/K selection theory, resource competition and community structure theory and the CSR model of life strategies.*

| Monod curves and other laboratory experiments | r/K selection theory | Resource competition theory | CSR model |
|---|---|---|---|
| Maximum specific growth rate $\mu_{max}$ | r | | $R_{max}$ |
| Intersection between specific growth and mortality rates | | R* | |
| Limiting nutrient concentration [S] | | Limiting resource concentration R | 1/Stress |
| Specific mortality rate | | Specific mortality rate | Disturbance |
| Yield for conversion of nutrient to biomass | Carrying capacity K | | Yield for conversion of nutrient to biomass |

In relation to Monod curves, the maximum specific growth rate (r) in r/K selection theory matches the maximum specific growth rate of a microbe in an isolated culture in the absence of nutrient limitations ($\mu_{max}$) (Figure 1, middle, and Table 1). It follows that r-strategists resemble copiotrophs and fast-growing microbes in that they present a higher value of r ($\mu_{max}$) than K-strategists, oligotrophs and slow-growing microbes (Figure 2 and Table 2). Although beyond the scope of a Monod plot, it should also be mentioned that the carrying capacity of an environment K is related to the nutrient-to-biomass yield (Table 1) and is higher in K-strategists relative to r-strategists (Table 2). In all, the partial correspondence between r/K selection theory and the Monod curves of copiotrophs/oligotrophs and fast versus slow-growing microbes suggests that quantitative ecological theory may help us understand these microbial dichotomies.

**Relation to resource competition and community structure theory**

The long term coexistence of a large number of species (Figure 1, left) cannot be explained using only the r/K categories. Tilman's resource competition and community structure theory was developed in the 1970s partly to address this issue (17) and builds on the fact that natural communities of organisms generally modify their environment by consuming resources to the point they limit their own growth. The theory models species coexistence by taking into account not only population growth, but also mortality and the consumption of limiting resources (Table 1). A main result is that as the population of an organism in the system grows, the amount of limiting resources (called R) trends down towards a limiting concentration, called R*, required to maintain a stable number of individuals. When R* is reached, consumption of the limiting resource equals resource renewal, and population growth evens out with mortality. Resource competition theory can be brought together with Monod curves by considering not only the specific growth rate but also the specific mortality rate, which we assume to be independent of limiting nutrient concentration (Figure 1, middle). It follows that R* is the limiting nutrient concentration for which the specific growth rate equals the specific mortality rate m (Table 1) and can be considered inversely proportional to the tolerance to nutritional stress. We find it useful to further analyze what happens at a limiting resource concentration equal to R*. At these conditions, we can substitute R* for [S] and m for μ in equation [1]. It follows that

$$\frac{R^*}{K_S} = \frac{m}{\mu_{max} - m} \qquad [2]$$

so that the ratio between R* and $K_s$ is determined by m and $\mu_{max}$. Thus, the R*/$K_s$ ratio between an important parameter of Monod curves and an indicator of ecological strategies can be determined using laboratory experiments. Recently, it was reported that the ratio $\mu_{max}$/m is approximately 7 for fifteen different microbes (25). As a consequence, the ratio R*/$K_s$ takes an approximately constant value close to 1/6. $K_s$ can be taken as proportional to R* and as inversely proportional to the tolerance to nutritional stress. In this case, tolerance to nutritional stress can be evaluated using the $K_s$ from a Monod curve.

Upon inspection of Figure 2, we can deduce that copiotrophs are expected to have a higher R* than oligotrophs and that fast-growing microbes are expected to have a lower R* than slow-growing microbes (Table 2), showing that resource competition theory can help discriminate models of microbial diversity. Previous experimental work on *Escherichia coli* and *Chelatobacter heintzii* indeed suggests that R* is a useful parameter for the study of resource competition (26).

More generally, resource competition theory shows that copiotrophic/oligotrophic and fast/slow-growing microbes consuming the same limiting resource cannot coexist at equilibrium (17), a result called the competitive exclusion principle (27). And yet, metagenomics and metatranscriptomics experiments show that they do (Figure 1, left), leading to an apparent paradox. One possible solution to the paradox is that copiotrophic/fast-growing microbes remain dormant unless limiting nutrient availability is high (6, 28). Community structure theory (17) provides a second solution to the paradox by showing that multiple species can coexist as long as there is differential resource utilization, that is, there are as many limiting resources as species and each species consumes more of the resource that most limits its own growth (29). This second solution seems to be relevant for microbial communities, since heterotrophic prokaryotes can grow on at least 107 different known carbon sources (1). Furthermore, meta-metabolomics of microbial communities identify up to 1000 different compounds that may sustain microbial growth (18). This suggests that, under favorable conditions, hundreds of different heterotrophic species should be able to coexist at equilibrium through nutritional niche differentiation. However, this leaves a gap between the number of species potentially sustained by nutritional differentiation and the number of observed species (Figure 1, left). Community structure theory can bridge this gap by showing how a spatially structured community at equilibrium allows for many more species than limiting resources (17), as can be readily observed in a Winogradsky column (3). Also, large communities may be stable due to syntrophy, where the side products of the metabolism of a microbe are essential for other microbes (29). Last, non-equilibrium conditions may further increase the number of coexisting species (29). The contrasting growth kinetics of copiotrophic/oligotrophic and fast/slow-growing microbes are likely to play a role in the dynamics of such large microbial communities.

**Relation to the CSR model of life strategies**

The CSR model was developed by Grime and coworkers in the 1970s to account for the diversity of life strategies in plants (30–32). The CSR model is similar to r/K selection theory in that it considers both a density-dependent control originated by limiting resources and a density-independent control originated by mortality. In contrast to r/K selection theory, in the CSR model limiting resources (also termed stress (30–32) or biomass restriction (33)) and mortality (also termed disturbance (30–32) or biomass destruction (33)), are taken into account as two separate environmental variables (Figure 3, left). This independence of limiting resources and mortality, i.e, of stress and disturbance, leads to the definition of three archetypal strategies (Figure 3, left). Competitive organisms (C strategists) have the highest fitness in environments with low stress/high resource availability and low disturbance/mortality. Stress-tolerant organisms (S strategists) have the highest fitness when disturbance/mortality is low but stress is high/resources are growth-limiting. Finally, ruderals (R strategists) have the highest fitness when there is low stress/an excess of resources but disturbance/mortality is high. Later work associated specific traits to each strategy in plants (30–32). The theory posits that due to trade-offs between the traits commonly associated with C, S and R strategies, an organism can not be simultaneously proficient at all strategies. A consequence of this is that if resources are scarce and mortality is high, no strategy leads to a viable population (Figure 3, left).

The CSR model has been adapted to describe microbial diversity, resulting in the "YAS triangle" (33, 34). Microbial high yield strategists (Y) are roughly analogous to competitor plants and are expected to maximize the fraction of nutrients that is allocated to the synthesis of biomass through the investment in anabolic pathways, i.e. to maximize yield. Yield is expected to present high values for Y(C) strategists, intermediate values for S strategists and low values for A(R) strategists (Table 2) (33). Microbes following the resource acquisition strategy (A) are analogous to ruderal plants and are expected to invest in the capture and assimilation of a wide range of substrates via motility, extracellular enzymes and catabolic pathways, i.e., to maximize the maximum potential specific growth rate (called $R_{max}$ in the CSR model). $R_{max}$ is expected to present high values for A(R) strategists, intermediate values for Y(C) strategists and low values for S strategists (Table 2) (33). Microbial S strategists resemble stress-tolerant plants and are expected to be proficient at maintaining existing biomass by making use of damage repair systems, cell protective structures and osmolytes, i.e., to maximize tolerance to nutritional stress. Stress tolerance is expected to present high values for S strategists, intermediate values for A(R) strategists and low values for Y(C) strategists (Table 2) (33). The values of stress and disturbance leading to no viable strategies are likely to be higher in microorganisms than in plants (35).

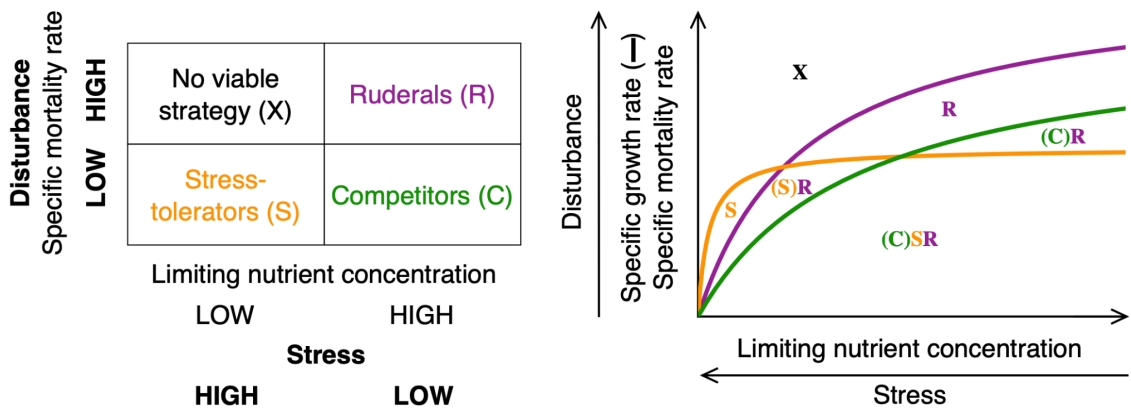

*Figure 3. Microbial dichotomies in the context of the CSR model of life strategies. Left: Different degrees of environmental stress (limiting nutrient concentration) and disturbance (specific mortality rate) lead to three viable strategies in the CSR model. Right: Monod curves for archetypal C strategists (green lines), S strategists (orange lines) and R strategists (purple lines). The areas delimited by the Monod curves are labeled with all the viable strategies under those conditions, with the strategy prevailing at high population densities between brackets. The position along the x axis indicates the level of a growth-limiting nutrient in the environment and is related to stress/biomass restriction and the position along the y axis is proportional to disturbance/biomass destruction/mortality and determines the minimal growth rate required to maintain a stable number of individuals.*

We present in the right panel of Figure 3 a simplified depiction of the relationship between growth rate and nutrient availability for C, S and R microbes in the context of Monod curves, using the Monod curves for archetypal strategists from Figure 2. We propose that the maximum specific growth rate of a microbe in an isolated culture ($\mu_{max}$) is a good proxy for $R_{max}$ because it represents an intrinsic maximum measured during balanced growth (Figure 1, middle, and Table 1). This is different from the effective specific growth rate in a natural community, which is modulated by the environment and by interactions with other species. We also argue that the concentration of a limiting nutrient in a Monod curve is inversely proportional to stress in the CSR model, while the specific mortality rate is analogous to disturbance in the CSR model (Figure 1, middle, and Table 1). Following the expectations from the CSR model (30–32), in the right panel of Figure 3 R strategists present the highest $R_{max}$, while S strategists grow fastest at low nutrient concentrations. C strategists are never the fastest to occupy the niche. However, they present the highest yield of nutrient-to-biomass conversion, which would lead them to prevail in a densely populated environment once the balanced growth characterized by the Monod curves plateaus.

The crossing of the three curves for C, S and R strategists defines several regions in the stress/disturbance parameter space (Figure 3, right). In order to deduce the strategies prevailing in each region of the parameter space, we should take into account both the specific growth rates for each of the three strategists (as indicated by the curves), the level of stress and the level of disturbance. For each point of the plot, the Monod curves above the point present a growth rate higher than the mortality rate and correspond to viable strategies, while the Monod curves below the point do not correspond to viable populations. Whenever multiple strategies are viable, we expect the strategy with the highest yield of nutrient-to-biomass conversion to prevail as high population densities are reached. In the upper left region of Figure 3 (right panel), the specific mortality rate is higher than the specific growth rates of all strategies and none of them leads to a viable population (this region is denoted by an X). In the S region, S strategists prevail because they are the only strategy with a specific growth rate higher than the specific mortality rate. Correspondingly, in the R region, R strategists prevail for the same reason. On the other hand, in the (S)R region both S and R strategists are viable, but S strategists are expected to prevail as high population densities are reached because of their higher yield of nutrient-to-biomass conversion (Table 2). Similarly, in the (C)R and (C)SR regions, even though multiple strategies are viable, C strategists are expected to take over the population as high population densities are reached because they present the highest yield of nutrient-to-biomass conversion (Table 2). In conclusion, analyzing archetypal Monod curves for the C, R and S strategists allows us to understand in quantitative terms how varying environmental conditions modulate the viability of microbial strategies.

In addition, the Monod curves of the CSR strategies provide a plausible explanation for the presence in the literature of the copiotroph/oligotroph and fast/slow-growing dichotomies (Figure 3, right). A comparison between an S strategist and a C or R strategist (orange versus green/purple lines) along a wide range of concentrations of a limiting nutrient would match the appearance of the copiotroph/oligotroph dichotomy (Figure 2, left), with C/R strategists being described as copiotrophs and S strategists as oligotrophs. Furthermore, a comparison between a C strategist and an R strategist (green versus purple lines) along a wide range of concentrations of a limiting nutrient would match the appearance of the fast/slow-growing microbes dichotomy (Figure 2, right), with R strategists being described as fast-growing and C strategists as slow-growing. Comparing microbes at narrow ranges of concentrations of a limiting nutrient would further complicate the picture. Thus, the difficulty of defining consistent dichotomies for the life strategies of microbes may be partly due to a more complex underlying repertoire of strategies related to the CSR model.

**An integrated view of microbial life strategies**

Is it possible to attain a coherent description of microbial life strategies that includes copiotrophs/oligotrophs, fast/slow growing microbes, r/K strategies and the CSR model? As mentioned along the previous sections, Table 2 summarizes the expectations for the values of the maximum specific growth rate ($\mu_{max}$, r, $R_{max}$), tolerance to nutritional stress (~1/R*) and the yield for conversion of nutrient to biomass (~carrying capacity K) derived from our analyses of Monod Curves and other considerations. Our proposal for an integration of these traits is described in the top panel of Figure 4 on the basis of Grime's CSR model. In this life strategy map, the trade-offs between stress tolerance and yield define a triangular shape for the space of viable strategies. The axes of variation for these three variables (Table 2) coincide with the sides of the triangle (Figure 4, top, continuous lines), with the archetypal strategies at the vertices. The growth, mortality and yield of each microbe in a particular environment determine its location in the triangle.

*Table 2. Microbial life strategies for the copiotrophic/oligotrophic, fast/slow-growing, r/K dichotomies and the CSR model.*

| Life strategies | Maximum specific growth rate ($\mu_{max}$, r, $R_{max}$) | Tolerance to nutritional stress (~$1/R^*$) | Yield (K, nutrient-to-biomass conversion) |
|---|---|---|---|
| **Copiotrophs/ oligotrophs** | Copiotrophs > Oligotrophs | Copiotrophs < Oligotrophs | Copiotrophs < Oligotrophs |
| **Fast/slow growing** | Fast-growing > Slow-growing | Fast-growing > Slow-growing | Fast-growing < Slow-growing |
| **r/K strategists** | r-strategists > K-strategists | Undetermined | r-strategists < K-strategists |
| **CSR model** | Ruderals > Competitors > Stress-tolerators | Stress-tolerators > Ruderals > Competitors | Competitors > Stress-tolerators > Ruderals |

We can locate other life strategy archetypes in a triangle using the data in Table 2 (Figure 4, top, dashed lines), with C, S and R strategists at the vertices. r/K strategists differ in $\mu_{max}$ and yield yet are undefined in terms of tolerance as formulated originally (21), placing this dichotomy along a height of the triangle. Thus, r-strategists (*sensu* r/K) are analogous to R strategists (*sensu* CSR), while K-strategists are in between C strategists and S strategists (Figure 4, top). In turn, fast/slow growing microbes differ in all three variables. Fast-growing microbes present high values of $\mu_{max}$ and tolerance and low values of yield relative to slow-growing microbes. The right side of the triangle separates fast and slow-growing microbes (Figure 4, top), since $\mu_{max}$ and tolerance increase towards the bottom right corner, while the yield decreases. This places slow-growing microbes together with C strategists and fast-growing microbes together with r-strategists and R strategists. Copiotrophs and oligotrophs also differ in all three variables. Copiotrophs present high values of $\mu_{max}$ and low values of tolerance and yield relative to oligotrophs (Figure 2, left) (10). This places the copiotrophic/oligotrophic dichotomy along a horizontal line inside of the triangle, since $\mu_{max}$ increases towards the right while tolerance and yield decrease. It follows that oligotrophs are in between S and C strategists, similar to K-strategists, and copiotrophs are in between R and C strategists (Figure 4, top). Overall, our approach of defining commonly discussed life strategies (Figure 4, top) with the help of Monod curves and the triangle yields a self-consistent picture.

We can group some, but not all, strategies from our version of the triangle. We display these associations in the bottom panel of Figure 4 in association with the Monod curves for archetypal microbes from the right panel of Figure 3. In this view, C strategists/slow-growing microbes are different names for the same archetype, as is the case for R strategists/fast-growing microbes/r-strategists. Oligotrophs and K-strategists may be considered a looser group since the former occupy a full side of the triangle and the latter should be located at the center of that side. Copiotrophs also occupy a full side of the triangle but can not be grouped with any other life strategy. These groupings are defined consistently but do not allow for a complete one-to-one correspondence of all the archetypes discussed here, as is also shown in the bottom panel of Figure 4.

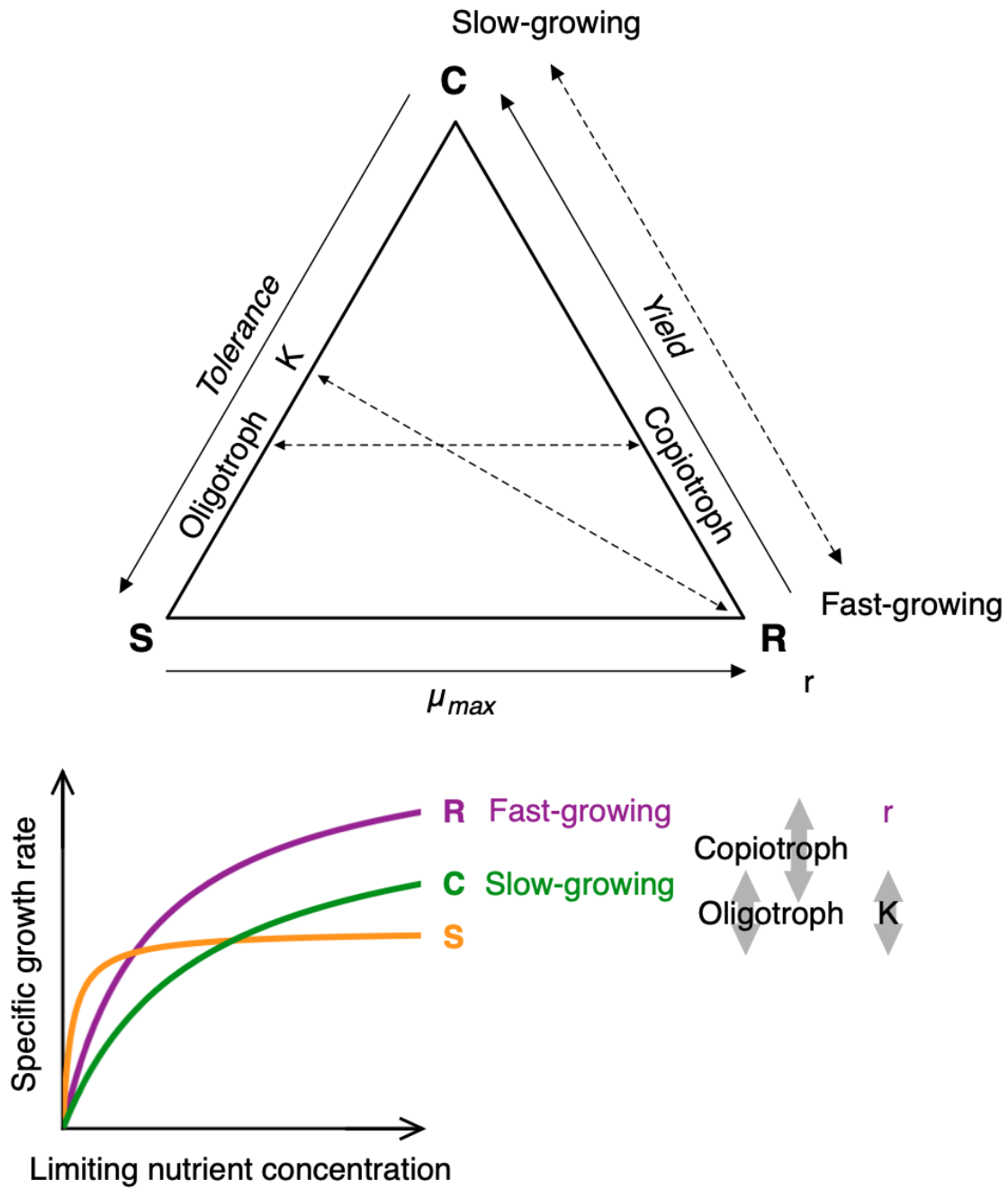

*Figure 4. An integrated view of microbial life strategies.* Top: The microbial life strategy triangle is defined in terms of $\mu_{max}$, stress tolerance and yield (continuous lines) following (30, 33, 34). The location of copiotrophs/oligotrophs, r/K strategists and fast/slow growing microbes (dashed lines) is defined using the traits in Table 2. Bottom: Microbial life strategies in relation to Monod curves and ecological theory. The Monod curves are the same as in Figure 3, left, while the groupings of archetypes are based on the top panel of this figure.

**Discussion**

Microbial growth has mostly been studied for a reduced number of model species in axenic cultures. Only relatively recently microbial ecology recovered some protagonism by probing the "biological dark matter" of non-culturable microbes. This was done thanks to housekeeping gene analysis technologies such as deep sequencing. With an ever increasing database of complete genomes, metagenomes and its associated environmental variables, the prediction of microbial physiology from genome sequences became achievable to a significant degree. The next step is to understand microbial populations and their dynamic successions based on environmental and metagenomic information. This will require context-dependent models of microbial behavior. Most models predicting microbial growth assume axenic conditions, and only recently some advances have been reported to incorporate genomic, physiological and ecological parameters into a unified theoretical framework of microbes in their environment. We wish to contribute to this discussion by combining classical ecology theories with classical bacterial physiology.

Coexisting copiotrophic/oligotrophic and fast/slow-growing microbes have long been put under the light of ecological theories, leading mainly to a discussion of the traits associated with the different life strategies (23, 33, 34). We explored quantitative relationships between the laboratory characterization of microbial growth using Monod curves and three classic ecology frameworks (Figure 1, middle). First, we found a straightforward correspondence between the $\mu_{max}$ parameter in Monod curves and the parameters r from r/K selection theory and $R_{max}$ from the CSR model. Second, the stress variable from Grime's CSR model is inversely proportional to the concentration of the limiting nutrient in the x axis of a Monod curve. Third, taking into account mortality rate allowed us to bring Tilman's R* and Grime's disturbance into Monod curves (Figure 1, middle). Fourth, the yield for nutrient-to-biomass conversion can not be extracted from Monod curves, yet can be readily measured in the laboratory (10) and is essential to the discussion of r/K selection theory and the CSR model. We propose that measurements of $\mu_{max}$ (Figure 1, right), R* and nutrient-to-biomass yield, together with a characterization of relevant environmental variables, should help us discuss microbes in their natural context.

We showed that Monod curves can reconcile the copiotroph/oligotroph and fast/slow-growing dichotomies in terms of an underlying tripartite CSR model (Figure 3, right). We take this to mean that these contrasting microbial dichotomies are not confusing nomenclature (5, 36) but *bona fide*, useful angles on a more complex model. We would like to underline that the time course of a microbial community greatly depends on the environmental conditions through their effects on the corresponding growth, mortality and yield of individual species. As a result, different microbes and physiological states may prevail in different circumstances. This fact, coupled to the "dormancy" capacity of microbes, can help explain the large microbial diversity that is sustained in nature.

We went on to use Monod curves (Figure 2) and other evidence to deduce values of $\mu_{max}$, yield and tolerance for the copiotrophic/oligotrophic, fast/slow-growing and r/K microbial dichotomies, as well as for the CSR life strategies (Table 2). The resulting differences allowed us to construct a revised microbial life strategy triangle that includes all these archetypes (Figure 4). Our view groups together fast-growing microbes with r-strategists (*sensu* r/K) and R strategists (*sensu* CSR), slow-growing microbes with C strategists (*sensu* CSR) and oligotrophs with K-strategists (Figure 4). On the other hand, oligotrophs and K-strategists may resemble S and C strategists depending on their tolerance values, and copiotrophs may resemble R and C strategists depending on their yield values (Figure 4). This supports the view that the copiotroph/oligotroph dichotomy does not entirely match neither the r/K dichotomy (12) nor the fast/slow-growing dichotomy. We are aware that there are as many definitions of copiotrophic/oligotrophic, r/K and fast/slow microbes as researchers, which would locate differently in the microbial life strategy triangle. To this, we would argue that locating individual microbes in the triangle via quantitative experimental measurements and the analysis of the resulting picture should help us resolve some of these ambiguities. We also note that the microbial life strategy triangle might be used to include other existing dichotomies (37) and to define additional, yet unnamed dichotomies along at least two sides and two heights of the triangle.

Our study also calls for a deeper characterization of experimental variables used in microbial cultures, such as pH, temperature or ionic strength. It is likely that such variables can not be considered as pure "stress variables" or "disturbance variables" but are better described quantitatively in relation to both stress and disturbance. For example, changes in the concentration of a growth-limiting nutrient, water activity or antibiotic concentration likely leads to physiological adaptation and changes in all three of growth, mortality and yield (38–40). As a consequence, alteration of environmental variables may change the position of a microbe in the triangle in any direction.

On the whole, we propose that quantitative studies are important for placing specific microbes in relation to life strategies (14). In the future, it may be interesting to integrate Monod curves, microbial life strategies and the different carbon sources for heterotrophic prokaryotes (Figure 1, left) into a theory for the diversity and structure of natural microbial communities. The integration of this information to genomic and metabolomic information may lead in the long term to design microbial communities for specific purposes.